\documentclass[]{spie}  %>>> use for US letter paper
\usepackage{amsmath,amsfonts,amssymb}
\usepackage{graphicx, placeins, epigraph}
\usepackage[colorlinks=true, allcolors=blue]{hyperref}
\title{Deep Snow: Synthesizing Remote Sensing Imagery with Generative Adversarial Nets}

\author[a]{Christopher X. Ren}
\author[a]{Amanda Ziemann}
\author[a]{James Theiler}
\author[b]{Alice M. S. Durieux}
\affil[a]{Intelligence and Space Research Division, Los Alamos National Laboratory, Los Alamos, NM}
\affil[b]{Descartes Labs, Inc, 100 N Guadalupe St, Santa Fe, NM }

\authorinfo{Send correspondence to Christopher X. Ren: \\E-mail: cren@lanl.gov}

\begin{document} 
\maketitle

\begin{abstract}
In this work we demonstrate that generative adversarial networks (GANs) can be used to generate realistic pervasive changes in RGB remote sensing imagery, even in an unpaired training setting. We investigate some transformation quality metrics based on deep embedding of the generated and real images which enable visualization and understanding of the training dynamics of the GAN, and provide a useful measure in terms of quantifying how distinguishable the generated images are from real images. We also identify some artifacts introduced by the GAN in the generated images, which are likely to contribute to the differences seen between the real and generated samples in the deep embedding feature space even in cases where the real and generated samples appear perceptually similar. 
\end{abstract}


\section{INTRODUCTION}

\epigraph{``Ceci n'est pas une pipe.''}{Ren\'e Magritte, \textit{The Treachery of Images}}

\label{sec:intro}  % \label{} allows reference to this section
The problem of mapping an image from one domain to a corresponding image in another domain is known as image-to-image translation. This type of task is analogous to language translation in the field of natural language processing (NLP): just as the same sentence may have representations in two different languages, a given image may have representations across two different domains. In this work, we attempt to generate synthetic color (three channel RGB) remote sensing imagery across seasonality domains. In particular, we focus on synthesizing snow cover. To this end, we utilize generative adversarial networks (GANs), a recent development in the field of deep generative modelling \cite{goodfellow2014generative}. The principal aim of this work is to investigate the viability of GANs as a method for the generation of synthetic remote sensing data which may be used to test a variety of change detection algorithms.

Interest within the field of remote sensing in utilizing GANs is growing, particularly in the realm of image-to-image translation, mostly stemming from the multi-modal nature of remote sensing imagery. Most of the efforts thus far in the field are directed at transforming data from different sensors into the same modality domain (usually multispectral, which is most easily interpreted)\cite{schmitt2018sen1, grohnfeldt2018conditional, ji2020sar, fuentes2019sar, schmitt2019sen12ms, toriya2019sar2opt}; these transformations are based on machine learning, not physical models, and so the transformed images comprise synthetic signals. One of the principal aims of these approaches is to take a modality unaffected by cloud cover, such as synthetic aperture radar (SAR), and learn a mapping to the multispectral domain which \emph{can} be hindered by cloud cover. Interestingly,  Mohajerani \textit{et al.} demonstrated the first use of an unpaired image-to-image translation framework in generating land cover changes (also snow to non-snow), but with the ultimate aim of preserving cloudy pixels for cloud segmentation purposes \cite{mohajerani2019cloudmaskgan} and limited to resolutions of 30 metres. In this work we demonstrate how an unmodified GAN-based framework can be utilized to generate high quality samples from Sentinel-2 imagery collected at a 10 metre resolution. Although previous work \cite{ren2019cycle} has investigated the practical applicability of this framework for anomalous change detection \cite{theiler2006proposed}, in this work we demonstrate metrics that enable the quantification of sample quality as well as evaluation of the viability of the transformed data in terms of generating synthetic imagery for potential applications.

\section{METHODS}

\subsection{Generative Adversarial Networks}
 
The framework of GANs consists of a so-called ``game'' between two adversaries. One is the generator, whose objective is to generate samples which originate from the same distribution as some training data. The other is the discriminator, whose objective is to determine whether samples are ``fake'' (generated by the generator) or ``real'' (originating from the training set) \cite{goodfellow2016nips}. Goodfellow \textit{et al.} liken the generator to a counterfeiter attempting to produce fake currency and distribute it undetected, and liken the discriminator to the police attempting to detect that fake currency \cite{goodfellow2014generative}. Competition in this process drives both the counterfeiter and the police to improve their methods of counterfeiting and detection, respectively, until the counterfeit currency is indistinguishable from the genuine article.

More formally: a prior over noise variables (the inputs) is defined as $p_{z}(z)$ in order to learn the distribution $p_{g}$ over data $\mathbf{x}$ (the training data) such that the mapping from the noise to the data space is $G(z;\theta_{g})$ where $G$ is a neural network with parameters $\theta_{g}$. A second neural network parametrized by $\theta_{d}$ is defined as $D(x, \theta_{d})$ and outputs a single scalar, the probability that a sample came from the training data rather than $p_{g}$. $D$ and $G$ are trained simultaneously in a two-player minimax game, such that $D$ attempts to maximize the probability of assigning correct labels to samples from the training data and those from $G$, whilst $G$ attempts to minimize $log(1 - D(G(z))$. The value function V of said minimax game is thus~\cite{goodfellow2014generative}:
\begin{equation}
\label{adversarial_loss}
    \min_{G}\max_{D} V(D, G) = \mathbb E_{x\sim p_{\mbox{\tiny d}}}[log(D(x)] + \mathbb E_{z\sim p_{z}}[1 - log(D(G(z)))].
\end{equation}

\subsection{Image-to-Image Translation}
The use of GANs as a potential general purpose solution to image-to-image translation problems was first highlighted by Isola \textit{et al.}  \cite{isola2017image}. In their framework, named \emph{pix2pix}, GANs were conditioned on input images in order to produce desired output images in a paired image setting. As such, the output produced by the GAN is evaluated both by the discriminator and by using an $L_{1}$ loss relative to the target.

In remote sensing imagery, it can often be difficult to consistently find corresponding pairs of imagery in the required domains at the same location due to cloud cover \cite{mohajerani2019cloudmaskgan}. To address this, we utilize an \emph{unpaired} image-to-image translation framework known as cycleGAN \cite{zhu2017unpaired}. This leverages a pair of GANs to learn \textit{two} transformations: one from domain X to Y, and one back from domain Y to X. These two GANs are bounded by the cycle-consistency loss: returning to our analogy of language translation, cycle-consistency ensures that a sentence translated from English to French, and then back to English is the same as the original sentence. In the case of cycleGAN, the two translators $G:X \rightarrow Y$ and $F:Y\rightarrow X$ should be inverses of one another and the translations are bijections, ensuring individual inputs and outputs can be connected in pairs in a valid manner \cite{zhu2017unpaired}.
Thus, given a pair of translators $G$ and $F$, the full objective is given by:
\begin{equation}
    L(G,F,D_{X}, D_{Y}) = L_{GAN}(G, D_{Y}, X, Y) + L_{GAN}(F, D_{X}, X, Y) + \lambda L_{cycle}(G, F)
\end{equation}
where $L_{GAN})$ is the adversarial loss function described by Eq.~\ref{adversarial_loss}; X and Y are the two domains between which we wish to learn a mapping; $D_{X}, D_{Y}$ are the discriminators paired with the generators $G, F$, respectively; and $L_{cycle}$ is the cycle-consistency loss given by:
\begin{equation}
    L_{cycle}(G, F) = \mathbb E_{x\sim p_{data}(x)}[||F(G(x)) - x||_{1}] + \mathbb E_{y\sim p_{data}(y)}[||G(F(y)) - y||_{1}].
\end{equation}

\subsection{Deep Embedding Quality Assessment}
\label{sec:deep quality}

The task of evaluating GANs is an area of active research, and a result of the fact that the optimization (driven by Eq.~\ref{adversarial_loss}) contains no explicit measure of sample quality other than the discriminator loss. Heusel \textit{et al.} introduced a measure known as the Fr\'echet Inception Distance, whereby samples are embedded in a specific layer of an Inception-v3 network \cite{szegedy2015going} pretrained on ImageNet in order to produce embedded features. Two multivariate Gaussians are then fit to these features and the Fr\'echet Distance \cite{dowson1982frechet}, or Wassterstein-2 distance \cite{vaserstein1969markov} between these two Gaussians ( fit to the real and fake embedded features) is calculated \cite{heusel2017gans} as:
\begin{equation}
    FID(x, g) = ||\mu_{x} - \mu_{g}||^{2}_{2} + Tr(\Sigma_{x} + \Sigma_{g} - 2(\Sigma_{x}\Sigma_{g}))^{\frac{1}{2}}
\end{equation}
where $\mu$ and $\Sigma$ denote the mean and covariance of the real (x) and generated (g) samples. In this work, we introduce a measure to evaluate the quality of samples from unpaired image-to-image translators such as cycleGANs. We utilize a similar methodology to Heusel \textit{et al.} in that we utilize a pretrained deep neural network to embed the samples. We choose the Resnet-50 architecture \cite{he2016deep} pretrained on ImageNet, as it has been shown to enable similarity searching in the embedding space for remote sensing imagery even without fine-tuning \cite{keisler2019visual}. Since the quality of the samples in this problem setting is determined not only by their realism but also by their ``closeness'' to the target domain, we evaluate the average cosine distances between each generated sample and each real sample in the Resnet-50 embedding space. The cosine Resnet distance (CRD) is then defined:
\begin{equation}
    CRD = \frac{1}{N} \sum\limits_{i=1}^N \frac{1}{M} \sum\limits_{j=1}^M G[i] \cdot X[j] 
\end{equation}
where $G[i]$ represents the $i$th generated sample embedded by the Resnet-50 network, and $X[j]$ represents the $j$th real sample also embedded by the Resnet-50 network. We also calculate the Fr\'echet Distance for the samples in the Resnet embedding space to provide a comparison between the Inception-v3 and Resnet features in terms of their applicability to remote sensing imagery.

\section{EXPERIMENT AND RESULTS}

\subsection{Generated Samples}

\begin{figure} [ht]
\begin{center}
\begin{tabular}{c} %% tabular useful for creating an array of images 
\includegraphics[width=10cm]{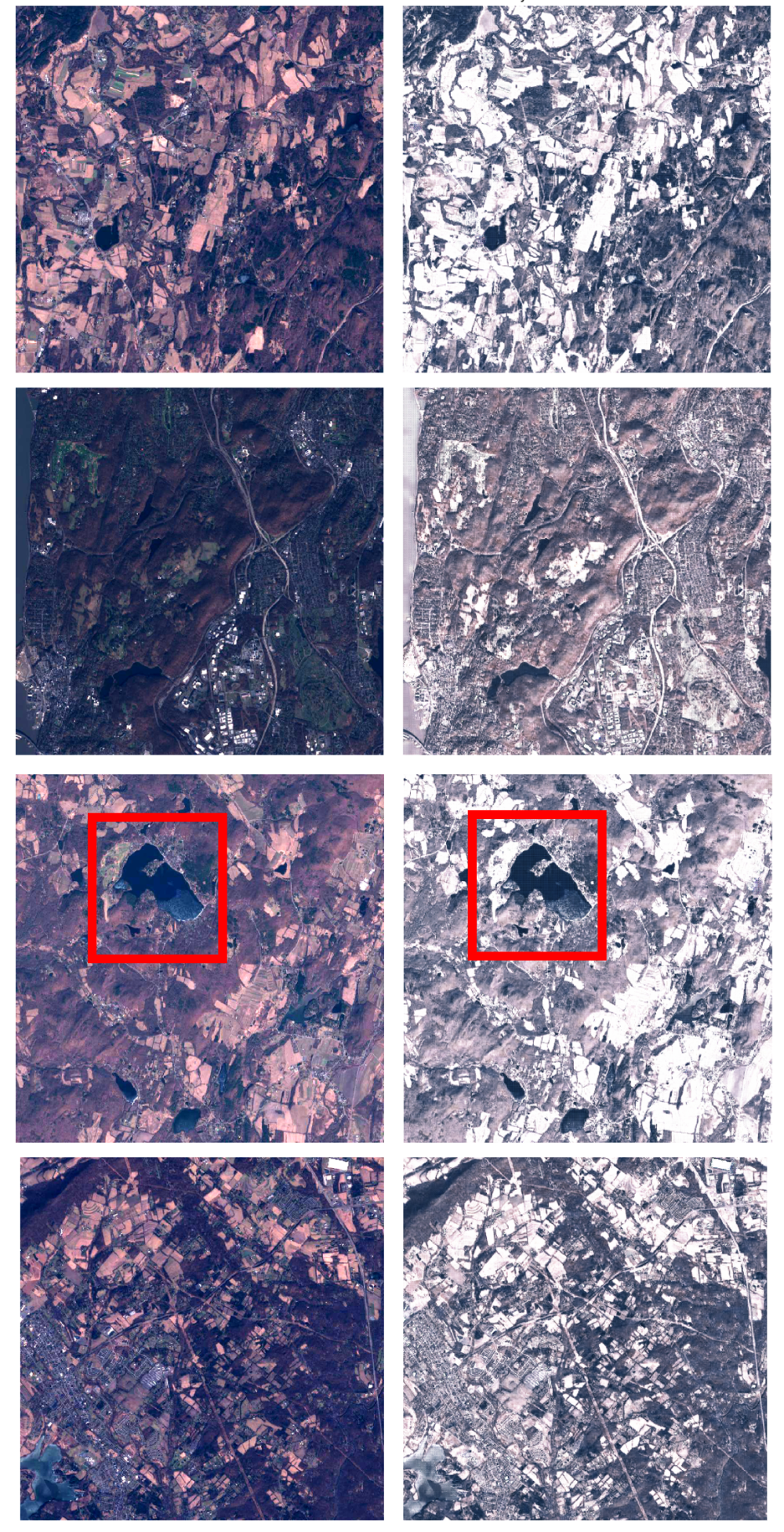}
\end{tabular}
\end{center}
\caption[example] 
%>>>> use \label inside caption to get Fig. number with \ref{}
{ \label{samples} 
Real Sentinel-2 RGB images (left column, with little or no snow), and the corresponding snow-transformed synthetic RGB images generated by the cycleGAN (right column). These transformed images attempt to realistically simulate what the original scenes would look like after snowfall.}
\end{figure}

Figure \ref{samples} shows some of the samples generated by the cycleGAN, where the input images are three channel RGB images from Sentinel-2, and the transformed images are those same scenes with realistic snow deposit. Qualitatively, we can see the high degree of realism in the generated snow. The cycleGAN is clearly able to learn the texture and colors of snow, and to apply them in a realistic manner while conserving the structure of the images. We note that the cycleGAN learns a perceptually realistic snow transformation for both rural and urban areas, colorizing roads, buildings and streets effectively.

% Note: If compiling with LaTeX+dvipdf, please ensure images generated from 
% other software packages have their bounding boxes set correctly.
\FloatBarrier

\begin{figure} [ht]
\begin{center}
\begin{tabular}{c} %% tabular useful for creating an array of images 
\includegraphics[width=13cm]{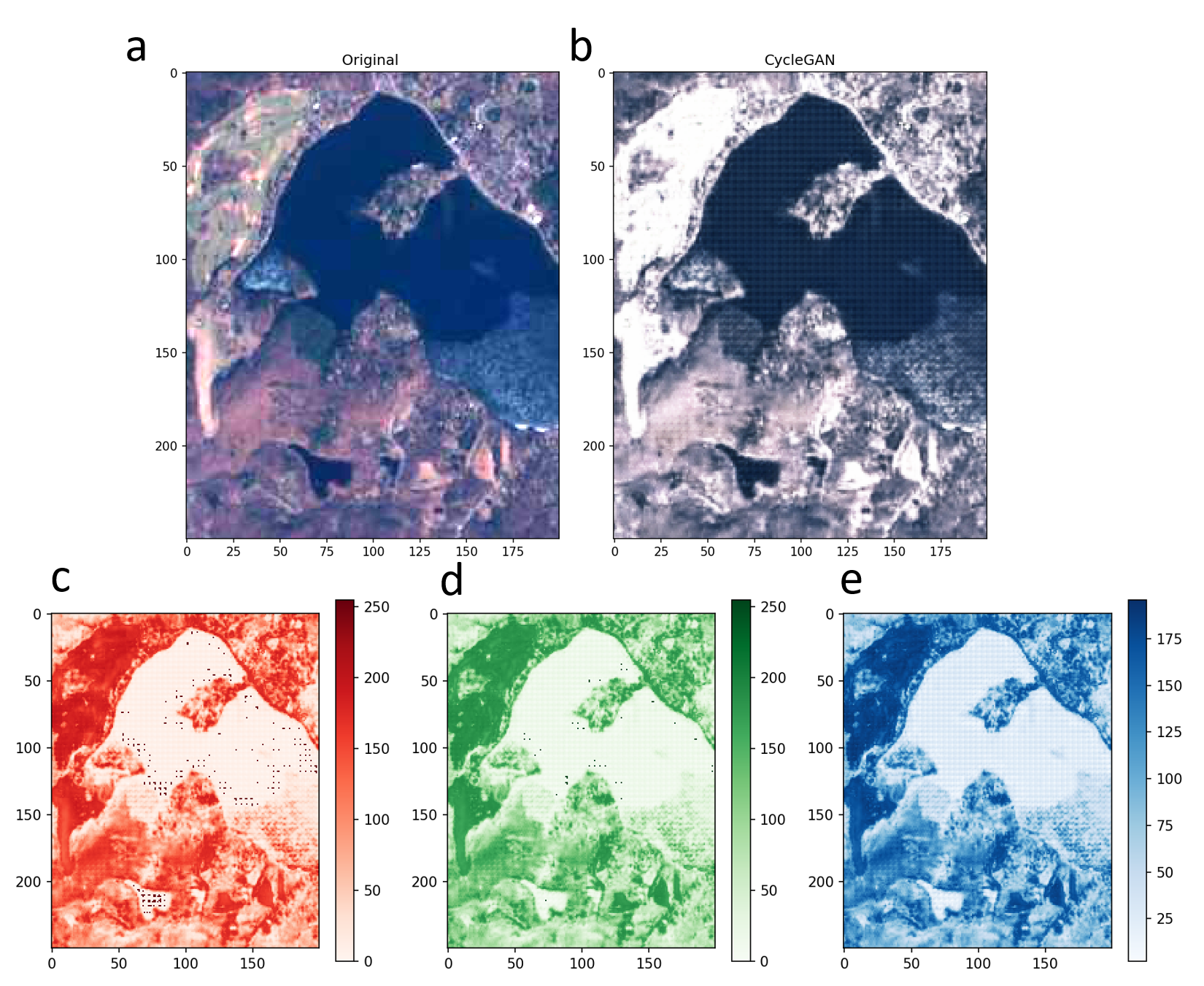}
\end{tabular}
\end{center}
\caption[example] 
%>>>> use \label inside caption to get Fig. number with \ref{}
{ \label{zoom} 
A closer look. (a) Original and (b) cycleGAN transformed region delineated by the red bounding box in Figure~\ref{samples}. (c) Red, (d) green, and (e) blue band L1 norms between (a) and (b).}
\end{figure}

Although the samples shown in Figure~\ref{samples} appear realistic, a closer look at some of the textures generated by the cycleGAN does reveal some issues. We show here a close up of the red bounding box in Figure~\ref{samples}. A common issue we observe in our dataset is the generation of unrealistic looking grid-like patterns, as shown in Figure~\ref{zoom}; in particular this is present over bodies of water. In order to further investigate the transformation learnt by the cycleGAN, we decompose the transformation into the red, green, and blue bands by simply taking the L1 norm of the bands between the original image and the transformed image. These band norms are shown in Figures~\ref{zoom}(c,d,e). We see here that not only does the cycleGAN introduce this grid-like artifact, but there also seems to be `noise'' that is imperceptible in the transformed image, but is evident in the speckle present in the L1 norm of the band intensities between the images, especially for the red band on the surface of the water. We hypothesize this effect occurs mostly over bodies of water as these are decidedly different from land in terms of texture in the images, and are much darker which can highlight low-amplitude effects; additionally, the cycleGAN sees few examples of bodies of water in both domains during the training process. 

\FloatBarrier
 
\subsection{Image Quality Assessment}

Thus far we have discussed the qualitative output of the cycleGAN. Currently there are no widely agreed upon and clearly-defined procedures for the evaluation of unpaired image-to-image translation frameworks. As mentioned in Section~\ref{sec:deep quality}, the evaluation of GAN output quality is an active area of research. The fact that our samples are unpaired further complicates the matter. For paired frameworks such as pix2pix, the image quality can be evaluated by comparing the generated image directly with the target; the inherently unpaired cycleGAN does not afford such a luxury. This introduces a further complexity: the unpaired nature of cycleGAN does not allow for the direct evaluation of a loss function, making it difficult to know when to terminate training (typically this would be determined by some stopping criterion based on the evaluation of a loss function). Although the unpaired nature of cycleGAN increases its generally applicability remote sensing problems, it does make it more difficult to assess image quality using traditional techniques.

\begin{figure} [ht]
\begin{center}
\begin{tabular}{c} %% tabular useful for creating an array of images 
\includegraphics[width=16cm]{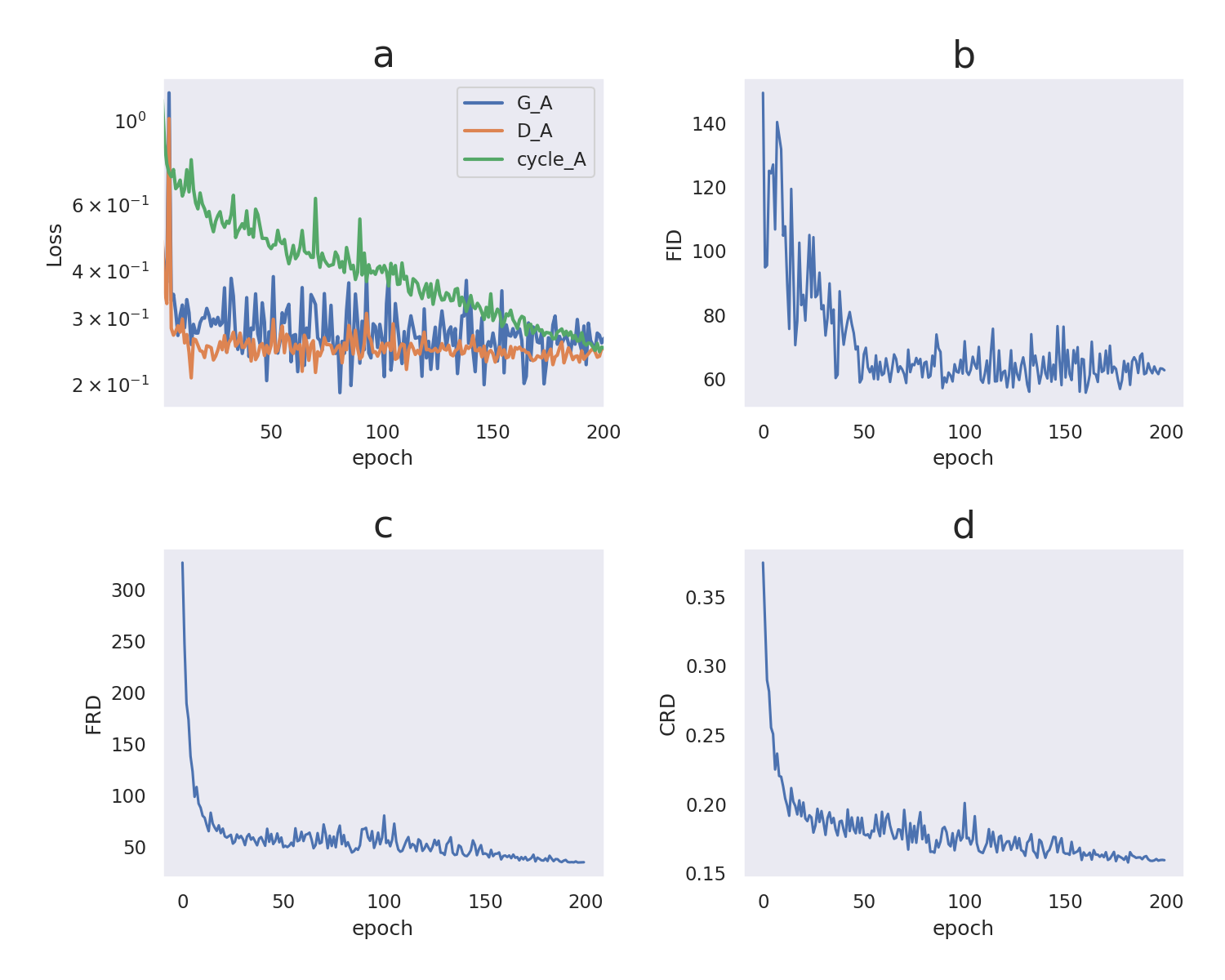}
\end{tabular}
\end{center}
\caption[example] 
%>>>> use \label inside caption to get Fig. number with \ref{}
{ \label{image_quality} 
CycleGAN evaluation metrics. (a) Generator, discriminator, and cycle-consistency losses (for the snow transformation only). b) Fr\'echet Inception Distance. (c) Fr\'echet Resnet Distance. (d) Cosine Resnet Distance.}
\end{figure}

In order to \emph{quantitatively} evaluate the progression of the cycleGAN, we apply the metrics described in Section~\ref{sec:deep quality} (FID, CRD and FRD) to data generated at each epoch of the training. This enables us to investigate the training dynamics of the cycleGAN, as shown in Figure~\ref{image_quality}. In Figure~\ref{image_quality}(a), we can see that only the cycle-consistency loss provides any indication that the cycleGAN output may be improving beyond the early stages of training (\~epochs 10-20), as both the generator and discriminator losses plateau early on. Similarly, the FID, shown in Figure~\ref{image_quality}(b), stops decreasing beyond epoch 50 and is additionally quite noisy throughout. Finally, Figures~\ref{image_quality}(c)~and~\ref{image_quality}(d) demonstrate that the embedding of remote sensing images using Resnet as opposed to Inception-v3 may capture more relevant features in the data, as we see our measures of transformation quality continuously decreasing throughout the training process in a much smoother manner compared to the FID.

\FloatBarrier

\begin{figure} [ht]
\begin{center}
\begin{tabular}{c} %% tabular useful for creating an array of images 
\includegraphics[width=14cm]{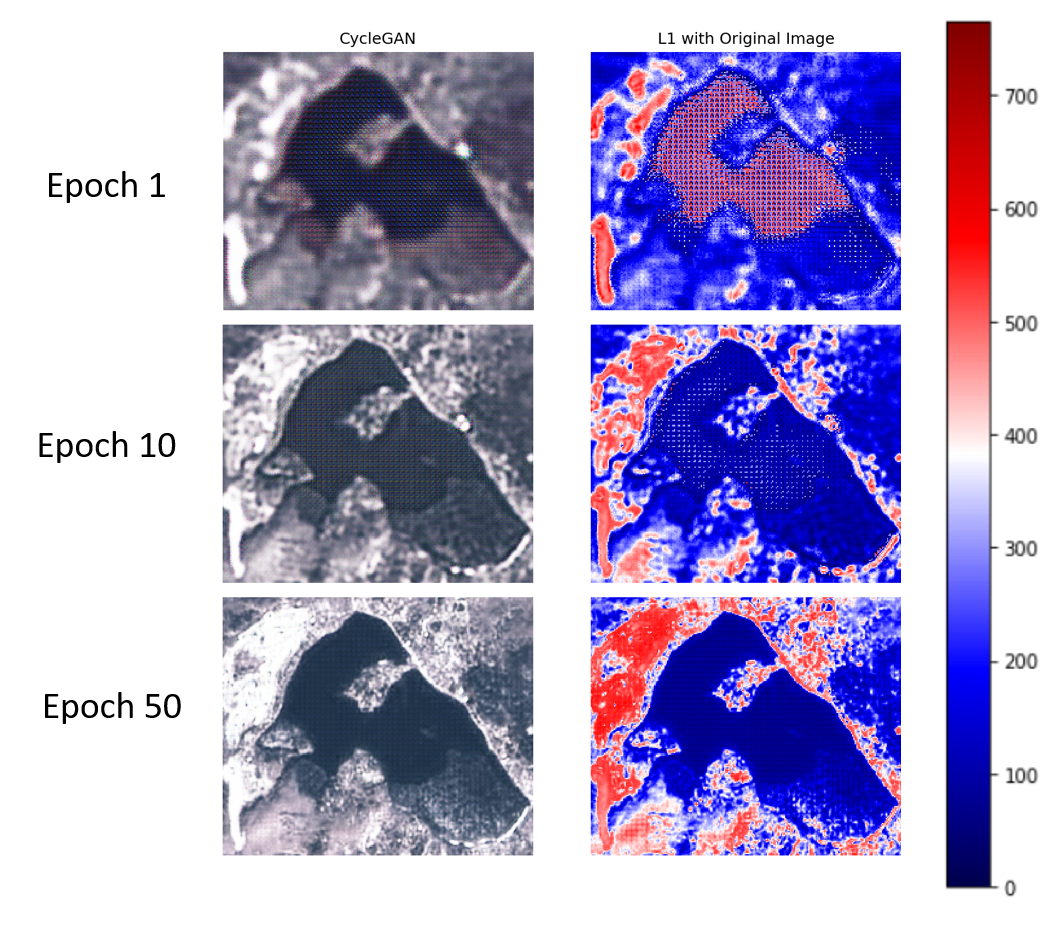}
\end{tabular}
\end{center}
\caption[example] 
%>>>> use \label inside caption to get Fig. number with \ref{}
{ \label{image_quality_2} 
Left column: samples taken from the same area as in Figure~\ref{zoom} at epochs 1, 10, and 50, respectively. Right column: Corresponding L1 of all three bands (RGB) between cycleGAN image and the original image to highlight the presence of artifacts over the body of water, and how these progressively disappear with training.}
\end{figure}

Figure~\ref{image_quality_2} again shows the same area as in Figure~\ref{zoom} for training epochs 1, 10 and 50. As the training progresses we see the image is gradually sharpened, an effect of the adversarial loss \cite{isola2017image}. We note that even in epoch 1, the ``colors'' of snow appear to have been learnt by the cycleGAN already, as the image is already vastly different to the original non-snowy image (shown in Figure~\ref{zoom}(a)). We show in the right-most column the L1 norm between all three bands of the cycleGAN transformed image in order to highlight artifacts generated by the GAN and show how these disappear as the training progresses.

\FloatBarrier

\subsection{Embedded Features}

We have shown in the previous section that quality metrics derived by embedding both real and cycleGAN generated images using a pre-trained Resnet-50 qualitatively correlate with perceptual image quality as evidenced by the decrease in the level of artifacts and blurriness of the cycleGAN generated images. As such, we now ask the question: \emph{based on this embedding, how convincing are the cycleGAN generated images relative to the real ones?}

\begin{figure} [ht]
\begin{center}
\begin{tabular}{c} %% tabular useful for creating an array of images 
\includegraphics[width=8cm]{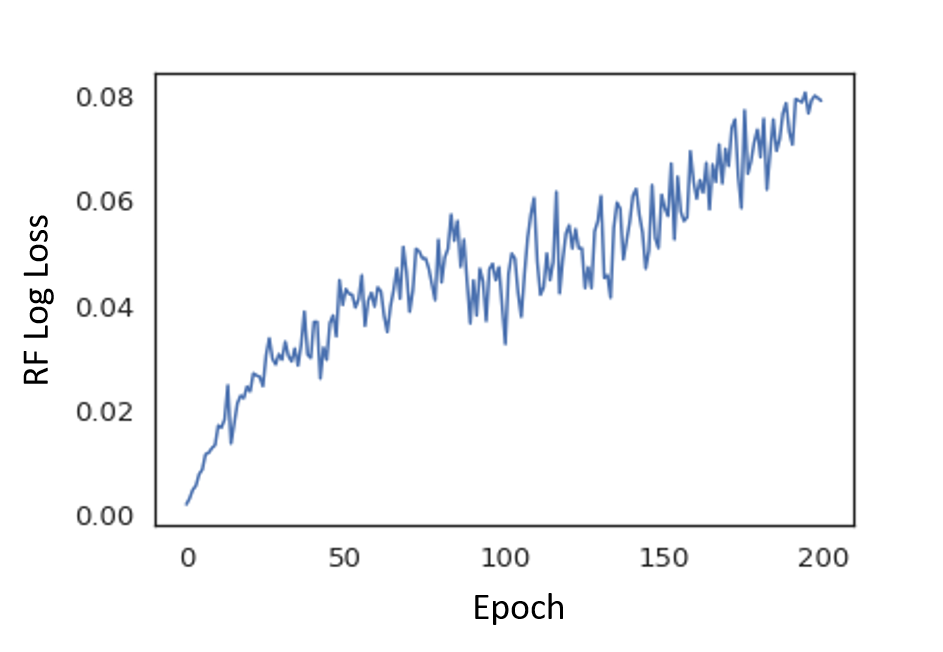}
\end{tabular}
\end{center}
\caption[example] 
%>>>> use \label inside caption to get Fig. number with \ref{}
{ \label{rf_loss} 
Random Forest classifier log loss for the task of classifying real and cycleGAN images based on their Resnet-50 features. The x-axis of this plot refers to the cycleGAN training epoch.}
\end{figure}

In a final test of the quality of the images generated by the cycleGAN, we train a random forest (RF) classifier to differentiate between real and cycleGAN generated images generated from each training epoch by embedding both sets of images using a pre-trained Resnet-50 and then performing 2-fold cross validation using an RF classifier averaged over 100 repeats. The averaged log loss of the RF classifier for generated data from each epoch is shown in Figure~\ref{rf_loss}. Although the generated image quality can be seen to improve in terms of the '`confusion,'' or increased log loss of the classifier in the task of classifying real and generated images, we note that even at epoch 200 the log loss remains low, indicating a good ability of the RF classifier to differentiate between real and synthetic images based on the Resnet-50 features. This may be due to artifacts such as those shown in Figure~\ref{zoom}.

\FloatBarrier

\begin{figure} [ht]
\begin{center}
\begin{tabular}{c} %% tabular useful for creating an array of images 
\includegraphics[width=11cm]{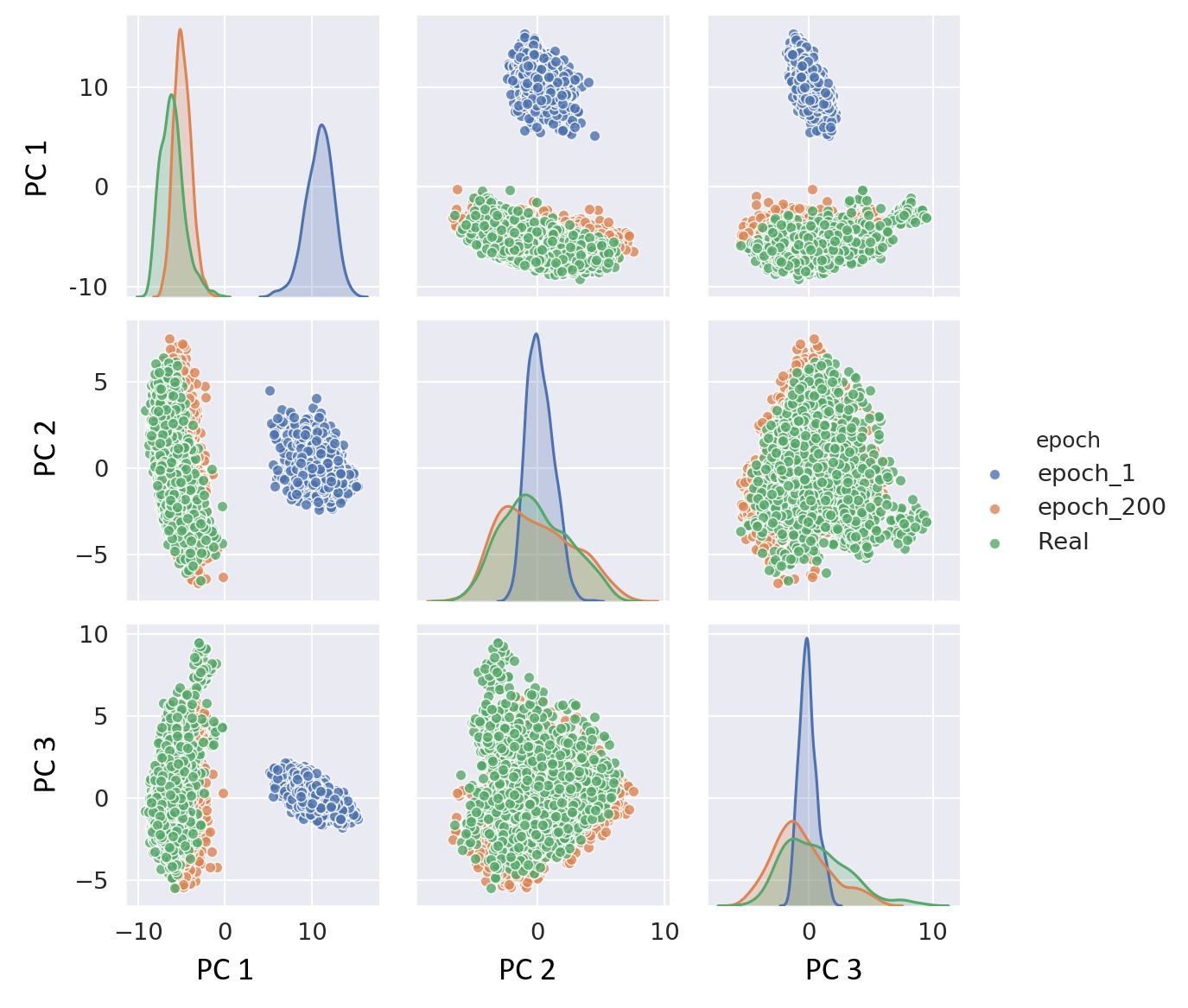}
\end{tabular}
\end{center}
\caption[example] 
%>>>> use \label inside caption to get Fig. number with \ref{}
{ \label{pair_plot} 
Pair plot showing the distribution of samples generated by the cycleGAN at epochs 1 (blue) and 200 (orange) relative to that of the real samples (green) for 3 principal components of the 2048-dimensional embedding space.}
\end{figure}

Finally, in order to further explore the behavior of the cycleGAN in terms of this embedding space, we perform principal component analysis (PCA) with 3 components on embedded features from epoch 1, 200, and the real samples, and visualize the results using a pair plot as shown in Figure~\ref{pair_plot}. Here, each off-diagonal plot illustrates a given principal component (PC) plotted against another, colored by the origin of the data (epoch 1, epoch 200, or real images). The diagonal plots show kernel density plots of the samples for each PC. We can see from these figures the ``distribution matching'' efforts of the cycleGAN in this embedding space: as training progresses, the distribution of the samples for these PCs in the embedding space goes from the blue distribution (epoch 1) to the orange one (epoch 200), modifying the distribution of the samples to match that of the real samples (green distribution).

\FloatBarrier

\section{Conclusions}

We have shown in this work that the cycleGAN framework can be utilized to learn realistic transformations of remote sensing imagery, with particular emphasis on transforming a non-snowy scene to a snow-covered one. We show that despite the apparent perceptual realism of the generated samples, a detailed look at some areas can reveal artifacts implanted by the transformation, which indicates that care must be taken when performing any sort of down-stream analysis on transformed samples. We introduce some methods to evaluate the quality of samples generated using unpaired image-to-image translation framework such as cycleGAN, which may help in guiding when training should be stopped. Although in this case we have only investigated modality-consistent image translations (i.e., Sentinel-2 RGB to Sentinel-2 RGB), we envision that image-to-image translation frameworks such as cycleGAN and pix2pix may be used in the future to generate synthetic datasets for the testing of algorithms designed to operate across multiple modalities~\cite{longbotham2012multi, ziemann2019multi, touati2019multimodal}. However, as demonstrated in this paper, careful analysis of potential artifacts introduced by these frameworks must be taken into account.

\section{Acknowledgements}

The research described in this paper was supported by the Los Alamos Laboratory Directed Research and Development (LDRD) program and the Center for Space and Earth Science (CSES). We also thank Descartes Labs for supporting streamlined imagery access and technical support. Finally, we thank Juston Moore, Jo\~ao, and Paulo M. for useful discussions.
% References
\bibliography{report} % bibliography data in report.bib
\bibliographystyle{spiebib} % makes bibtex use spiebib.bst

\end{document}